\documentclass[12pt]{article}
\usepackage{amssymb,amsmath,epsfig}
\allowdisplaybreaks

\begin{document}
\title{\bf Impact of Charge on Traversable Wormhole Solutions in f(R,T) Theory}
\author{M. Sharif \thanks {msharif.math@pu.edu.pk} and Arooj Fatima
\thanks{arooj3740@gmail.com}\\
Department of Mathematics and Statistics,\\
The University of Lahore, 1-KM Defence Road Lahore, Pakistan.}
\date{}
\maketitle

\begin{abstract}
This paper examines the effects of charge on traversable wormhole
structure in $f(R,T)$ theory. For this purpose, we use the embedding
class-I approach to build a wormhole shape function from the static
spherically symmetric spacetime. The developed shape function
satisfies all the required conditions and connects two
asymptotically flat regions of spacetime. We consider different
models of this modified theory to examine the traversable wormhole
solutions through null energy condition and also check their stable
state. We conclude that viable and stable wormhole solutions are
obtained under the influence of charge in this gravitational theory.
\end{abstract}

\section{Introduction}

Different cosmic observations such as supernovae type 1a, the
large-scale structures and the cosmic microwave background
radiations reveal that our universe is in accelerated expansion
phase \cite{1}. Researchers claim that this expansion is the result
of some mysterious force known as dark energy. The lambda cold dark
matter model based on general relativity (GR) is the first model
which provides a mathematical framework for understanding the
effects of dark energy on the spacetime. However, this model has
fine-tuning and coincidence problems. Different modified theories
have been developed to resolve these issues. The $f(R)$ gravity is
the first and simplest modification which is obtained by
incorporating the Ricci scalar $(R)$ with its generic function in
the Einstein-Hilbert action. The important literature is available
for understanding various characteristics of this modified theory
\cite{2}-\cite{5}. Harko et al \cite{6} introduced $f(R,T)$ theory
by incorporating the trace of energy-momentum tensor (EMT) in the
functional action of $f(R)$ gravity. This is one of the modified
theories that has been established to study the impact of
curvature-matter interaction on cosmic structures
\cite{7}-\cite{16}. Sharif and Waseem \cite{17} used the Krori-Barua
solutions and examined the viability of anisotropic quark stars in
this theory.

The study of hypothetical structures in the universe such as
wormhole (WH) is a fascinating area of research. The term WH refers
to a non-singular solution of the field equations that creates a
shortcut among two distinct areas of the universe. Wormholes are
thought to be formed by a process called ``gravitational collapse''.
This occurs when a massive object collapses due to the force of
gravity. The resulting spacetime distortion may create a bridge or
WH between two distant points in the universe. An intra-universe WH
connects two distant points within the same universe and an
inter-universe WH connects two different universes. Flamm \cite{19}
was the first who constructed the isometric embedding of the
Schwarzschild solution, which is considered to be an early precursor
to the concept of WHs. Later, Einstein and Rosen \cite{20}
investigated that the spacetime can be connected by a tunnel-like
structure, allowing for a shortcut or bridge between two distant
regions of spacetime, called \emph{Einstein-Rosen bridge}. Wheeler
\cite{20a} showed that Schwarzschild WH solutions are
non-traversable because the throat of the WH can change its size and
shape. Ellis \cite{21} invented the term ``Drainhole'' to describe a
WH.

In 1988, Morris and Thorne \cite{22} published a paper in which they
proposed the concept of a traversable WH. A WH is ``traversable'' if
it allows the matter to travel without getting destroyed by the
extreme gravitational forces present inside it. They proposed that a
WH can be traversable if it is stabilized by a form of exotic matter
with negative energy density. This exotic matter would need to have
specific properties such as negative pressure to counteract the
gravitational pull otherwise causes the WH to collapse. Exotic
matter can be used to control the size and shape of the WH throat,
preventing it from closing off too quickly. One proposal for
confining exotic matter in the interior space of WH is through the
use of ``matching conditions'' \cite{23}. These conditions are
mathematical equations that define how the geometry of one region of
a WH match with the geometry of another region for traversable and
stable WH structure. Spherical traversable WHs are most commonly
studied and can only exist in the presence of exotic matter
\cite{24}. Additionally, phantom energy has been used to construct
static spherical WHs \cite{25,26}. It is believed that cylindrical
WHs may be able to exist under slightly different conditions as
compared to spherically symmetric WHs \cite{27}-\cite{29}. Gibbons
and Volkov \cite{30} discussed how Einstein-Rosen and Flamm's
solutions relate to each other.

Shape functions are important in determining the properties and
behavior of traversable WHs. It is a mathematical function that
describes the spatial geometry of the WH, specifically the radius of
throat as a function of the radial coordinate. Different shape
functions can be used to model different types of WHs. For example,
the Morris-Thorne shape function is commonly used to model WHs with
spherical symmetry \cite{31}. The choice of shape function can
greatly affect the properties of WH such as stability,
traversability and the required amount of exotic matter to keep the
WH throat open. Sharif and Fatima \cite{32} used two different shape
functions to investigate the viability of non-static conformal WHs.
Cataldo et al \cite {33} studied static traversable WH solutions by
constructing a shape function that connects two asymptotically
non-flat regions of spacetime. In the framework of $f(R)$ gravity,
Godani and Samanta \cite{34} used two different shape functions and
introduced new WH structures. Sharif and Gul \cite{35} used Noether
symmetry technique to check the viability and stability of WH
geometry corresponding to different redshift and shape functions in
$f(R,T^{2})$ theory, where $T^{2}$ is the self-contraction of EMT.

The effect of electric charge can significantly impact the geometry
and stability of WHs. In particular, the electric charge can create
a repulsive force that pushes the walls of WH apart making the
throat of WH wider. This can affect the stability of WH as it may
require a greater amount of energy to keep the throat open. There
has been several works exploring the influence of charge on the
structure of WH. Reissner and Nordstr$\ddot{o}$m presented the first
static charged solution of the Einstein-Maxwell field equations,
which describe the gravitational and electromagnetic fields in GR.
Esculpi and Aloma \cite{37} studied the impact of anisotropy on
charged compact objects by using a linear equation of state. Sharif
and Mumtaz constructed charged thin-shell WHs using Visser cut and
paste approach \cite{38}. Moraes et al \cite{39} constructed charged
WHs in the context of curvature matter coupled gravity. Sharif and
Naz \cite{40} studied the dynamics of charged cylindrical collapse
in $f(G)$ theory ($G$ is the Gauss-Bonnet invariant) and found that
Gauss-Bonnet terms and charge prevent gravitational collapse. Sharif
and Javed \cite{41} used the cut and paste approach to examine the
influence of charge and Weyl coupling parameters on thin-shell WHs.

The scientific community has shown great interest in studying the WH
structures in modified theories. Elizalde and Khurshudyan \cite{42}
used the barotropic equation of state to examine the viability and
stability of WH geometry in curvature-matter coupled gravity. Sharif
and Shahid \cite{43} used the Noether symmetry technique to study
the WH solutions with isotropic matter configuration in $f(G,T)$
gravity. Shamir and Fayyaz \cite{44} built a WH shape function using
Karmarkar condition in the modified $f(R)$ theory. In the same
theory, Mishra et al \cite{45} found traversable WH solutions using
three different models of $f(R)$ gravity. Shamir et al \cite{46}
analyzed WH solutions with anisotropic matter configuration in
$f(R)$ gravity. Recently, we have employed the Karmarkar condition
and investigated the viable WH structure with anisotropic matter
configuration in $f(R,T)$ theory \cite{47}.

This paper analyzes the impact of charge on traversable WH solutions
using the Karmarkar condition in $f(R,T)$ theory. We have arranged
the paper as follows. We construct the shape function admitting
Karmarkar condition in section \textbf{2}. In section \textbf{3}, we
use charged anisotropic matter configuration to construct the field
equations and examine the nature of null energy condition by
considering two different models of this gravity. The stable state
of the resulting WH solutions is checked in section \textbf{4}. In
the last section we compile our obtained results.

\section{Karmarkar Condition}

Here, we employ the Karmarkar condition to develop the WH shape
function which describes the WH geometry. In this perspective, we
assume static spherically symmetric spacetime as
\begin{equation}\label{1}
ds^{2}=e^{\lambda(r)}dt^{2}-e^{\psi(r)}dr^{2}-r^{2}(d\theta^{2}
+\sin^{2}\theta d\phi^{2}).
\end{equation}
The non-zero components of the curvature tensor for the above
spacetime are given as
\begin{eqnarray}\nonumber
R_{1212}&=&\frac{e^{\lambda}(2\lambda''+\lambda'^{2}-\lambda'\psi')}{4},
\quad R_{3434}=\frac{r^{2}\sin^{2}\theta(e^{\psi}-1)}{e^{\psi}},
\\\nonumber
R_{1414}&=&\frac{r\sin^{2}\theta\lambda'e^{\lambda-\psi}}{2}, \quad
R_{2323}=\frac{r\psi'}{2}, \quad R_{1334}=R_{1224}\sin^{2}\theta,
\end{eqnarray}
where $R_{1224}=0$. These components fulfill the well-known
Karmarkar condition as
\begin{eqnarray}\label{2}
R_{1414}&=&\frac{R_{1212}R_{3434}+R_{1224}R_{1334}}{R_{2323}},\quad
R_{2323}\neq0.
\end{eqnarray}
Embedding class-I is the type of spacetime that fulfills the
Karmarkar condition. By substituting the values of non-zero Riemann
components in the Karmarkar condition, we get
\begin{equation}\nonumber
\frac{\lambda'\psi'}{1-e^{\psi}}=\lambda'\psi'-2\lambda''-\lambda'^{2},
\quad e^{\psi}\neq1.
\end{equation}
We solve the above equation and obtain the solution as
\begin{equation}\label{3}
e^{\psi}=1+C e^{\lambda}\lambda'^{2},
\end{equation}
here $C$ is an arbitrary constant.

In order to construct the WH shape function, we take the
Morris-Thorne metric as
\begin{equation}\label{4}
ds^{2}=e^{\lambda(r)}dt^{2}-\frac{1}{1-\frac{d(r)}{r}}dr^{2}
-r^{2}d\theta^{2}-r^{2}\sin\theta d\phi^{2}.
\end{equation}
Here, $d(r)$ is the shape function and $\lambda(r)$ is the redshift
function which is defined as $\lambda(r)=\frac{-2\epsilon}{r}$,
where $\epsilon$ is an arbitrary constant and, when
$r\rightarrow\infty$, $\lambda(r)\rightarrow0$ \cite{48}. From
Eqs.(\ref{1}) and (\ref{4}), we have
\begin{equation}\label{5}
\psi(r)=\ln\bigg[\frac{r}{r-d(r)}\bigg].
\end{equation}
We use Eqs.(\ref{3}) and (\ref{5}) to get the shape function as
\begin{equation}\label{6}
d(r)=r-\frac{r^{5}}{r^{4}+4\epsilon^{2}Ce^{\frac{-2\epsilon}{r}}}.
\end{equation}
The following conditions of shape function must be satisfied for a
viable WH structure \cite{22}.
\begin{enumerate}
\item
$d(r)<r$,
\item
$d(r)-r=0$ at $r=b$,
\item
$\frac{d(r)-rd'(r)}{d^{2}(r)}>0$ at $r=b$,
\item
$d'(r)<1$,
\item
The condition $\frac{d(r)}{r}\rightarrow0$ should be satisfied when
$r\rightarrow\infty$,
\end{enumerate}
where $b$ is known as WH throat radius. At WH throat, Eq.(\ref{6})
gives a trivial solution such that $d(b)-b=0$ at $b=0$. We add a
free parameter $(A)$ in Eq.(\ref{6}) to get a non-trivial solution
as
\begin{eqnarray}\label{7}
d(r)=r-\frac{r^{5}}{r^{4}+4\epsilon^{2}Ce^{\frac{-2\epsilon}{r}}}+A.
\end{eqnarray}

For a viable WH geometry, these conditions must be fulfilled. These
conditions are satisfied for $0<A<b$, otherwise the required
conditions are not satisfied and one cannot obtain the viable WH
structure. By using condition 2 in the above equation, we get
$C=\frac{b^{4}(b-A)}{4b\epsilon^{2}e^{\frac{-2\epsilon^{2}}{b}}}$.
Inserting this value in Eq.(\ref{7}), we get the shape function as
\begin{eqnarray}\label{8}
d(r)=r-\frac{r^{5}}{r^{4}+b^{4}(\frac{b}{A}-1)}+A.
\end{eqnarray}
We can see that our developed shape function also satisfies the
other conditions and hence asymptotically flat traversable WH
geometry is obtained. For our convenience, we assume $b=2$ and
$\epsilon=-1$ to analyze the graphical nature of the WH shape
function. Figure \textbf{1} indicates that for different values of
$A$, all conditions meet the required criteria. The values of
$A=1.9,1.8,1.7,1.6$ represent black, green, red and yellow colors,
respectively.
\begin{figure}
\epsfig{file=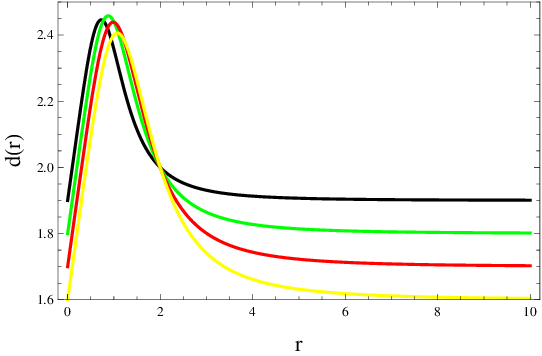,width=.5\linewidth}
\epsfig{file=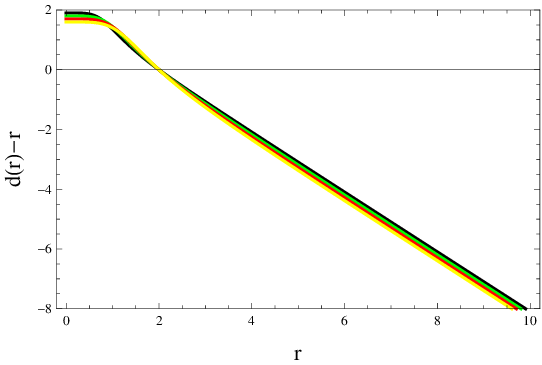,width=.5\linewidth}
\epsfig{file=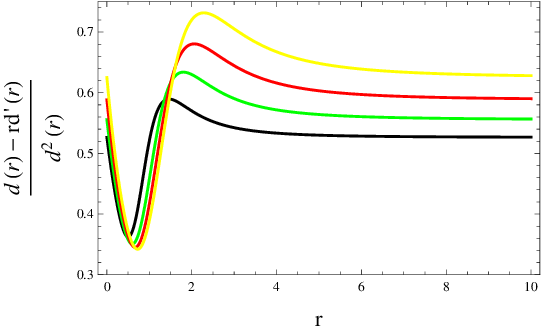,width=.5\linewidth}
\epsfig{file=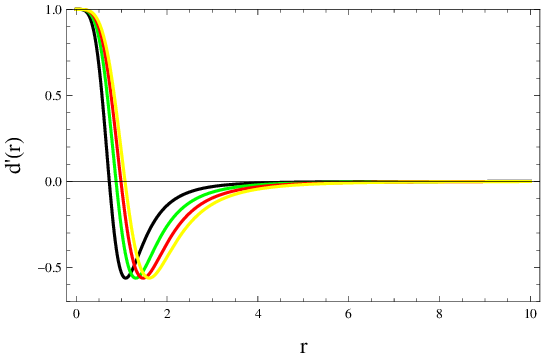,width=.5\linewidth}\center
\epsfig{file=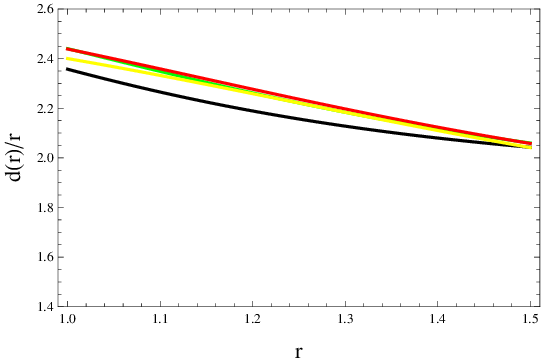,width=.5\linewidth}\caption{Plots of shape
function conditions with respect to radial coordinate.}
\end{figure}

\section{$f(R,T)$ Gravity}

The integral action for $f(R,T)$ theory is given as \cite{49}
\begin{equation}\label{9}
I=\int dx^{4}\sqrt{-g}\bigg[\frac{f(R,T)}{2k}+L_{m}+L_{e}\bigg],
\end{equation}
where $L_{m}$ is the matter-Lagrangian, $g$ is the determinant of
the metric tensor and $L_{e}=sF_{\chi\eta}F^{\chi \eta}$, where $s$
is an arbitrary constant. Here $F_{\chi
\eta}=\phi_{\eta,\chi}-\phi_{\chi,\eta}$ is the electromagnetic
field tensor and $\phi_{\chi}$ is the four potential. The
corresponding field equations are
\begin{equation}\label{10}
R_{\chi\eta}f_{R}-\frac{1}{2}g_{\chi\eta}f
-(\nabla_\chi\nabla_\eta-g_{\chi\eta}\Box)f_{_{R}}+f_{T}\Theta
_{\chi\eta}+f_{T}T_{\chi\eta}=T_{\chi\eta}+E_{\chi \eta}.
\end{equation}
Here, $f_{R}=\frac{\partial f}{\partial R}$ and
$f_{T}=\frac{\partial f}{\partial T}$. The equation of
$\Theta_{\chi\eta}$ is given as
\begin{equation}\label{11}
\Theta
_{\chi\eta}=-2T_{\chi\eta}+g_{\chi\eta}L_{m}-2g^{\varpi\sigma}
\frac{\partial^{2}L_{m}}{\partial g^{\chi\eta}\partial
g^{\varpi\sigma}},
\end{equation}
and EMT of electric field is expressed as
\begin{equation}\label{12}
E_{\chi\eta}=\frac{1}{4\pi}\bigg(\frac{F^{\varpi\sigma}F_{\varpi\sigma}
g_{\chi\eta}}{4}-F^{\varpi}_{\chi}F_{\eta\varpi}\bigg).
\end{equation}
We assume that the WH interior geometry is filled with anisotropic
fluid, given as
\begin{equation}\label{13}
T_{\chi\eta}=(P_{r}-P_{t})U_{\chi}U_{\eta}-P_{t}g_{\chi
\eta}+(\rho+P_{t})V_{\chi}V_{\eta},
\end{equation}
where $\rho$ represents the energy density, $V_{\chi}$ is the
four-velocity, $U_{\chi}$ defines four-vector, $P_{r}$, $P_{t}$ are
radial and tangential pressure, respectively. We consider
$L_{m}=-\frac{1}{4}F_{\chi\eta}F^{\chi\eta}$ which leads to
$\frac{\partial^{2}L_{m}}{\partial g^{\chi\eta}\partial
g^{\varpi\sigma}}=-\frac{1}{2}F_{\chi\varpi}F_{\eta\sigma}$
\cite{50}. Thus, Eq.(\ref{11}) becomes
\begin{equation}\nonumber
\Theta
_{\chi\eta}=-2T_{\chi\eta}+g_{\chi\eta}L_{m}+2g^{\varpi\sigma}
F_{\chi\varpi}F_{\eta\sigma}.
\end{equation}
We can rewrite Eq.(\ref{10}) as
\begin{equation}\label{14}
G_{\chi\eta}=\frac{1}{f_{R}}[T_{\chi\eta}+E_{\chi\eta}+T^{d}_{\chi\eta}],
\end{equation}
where $T^{d}_{\chi\eta}$ are the additional effects of $f(R,T)$
theory named as dark source terms, expressed as
\begin{equation}\label{15}
T^{d}_{\chi\eta}=\bigg[\frac{1}{2}(f-Rf_{R})g_{\chi\eta}-(g_{\chi\eta}
\Box-\nabla_\chi\nabla_\eta)f_{R}-(T_{\chi\eta}+\Theta_{\chi\eta})f_{T}\bigg].
\end{equation}

The Maxwell field equations are defined as
\begin{eqnarray}\label{16}
F_{\chi\eta;\varpi}=0, \quad F^{\chi\eta} _{;\sigma}=4\pi J^{\chi},
\end{eqnarray}
where $J^{\chi}$ is defined as four current. In comoving
coordinates, both $\phi^{\chi}$ and $J^{\chi}$ fulfill the following
relations
\begin{eqnarray}\label{17}
\phi^{\chi}=\phi(r)\delta^{\chi}_{0}, \quad
J^{\chi}=\varrho(r)V^{\chi},
\end{eqnarray}
where $\varrho$ is the charge density. The resulting electromagnetic
field equation is
\begin{equation}\label{18}
\phi''+\bigg(\frac{2}{r}-\frac{\lambda'}{2}-\frac{\psi'}{2}\bigg)\phi'=4\pi
\varrho(r)e^{\frac{\lambda}{2}+\psi}.
\end{equation}
Here, prime is the derivative corresponding to radial coordinate.
Integrating Eq.(\ref{18}), we get
\begin{eqnarray}\label{19}
\phi'=\frac{qe^{\frac{\lambda+\psi}{2}}}{r^{2}}, \quad
q(r)=4\pi\int_{0}^{r}\varrho(r)r^{2}e^{\frac{\psi}{2}}dr, \quad
E=\frac{q}{4\pi r^{2}},
\end{eqnarray}
where $E$ is the charge intensity and $q$ represents the charge
inside the interior of WH. By solving Eqs.(\ref{1}) and (\ref{14}),
we obtain the field equations of charged anisotropic spherical
system as
\begin{eqnarray}\label{20}
\rho&=&\frac{1}{e^{\psi}}\bigg[-e^{\psi}\big(\frac{f}{2}+\frac{q^{2}}{8\pi
r^{4}}+\frac{q^{2}}{2r^{4}}f_{T}\big)+\big(\frac{\lambda'}{r}
-\frac{\lambda'\psi'}{4}+\frac{\lambda''}{2}+\frac{\lambda'^{2}}{4}\big)f_{R}
\\\nonumber
&+&\big(\frac{\psi'}{2}-\frac{2}{r}\big)f'_{R}-f''_{R}\bigg],
\\\nonumber
P_{r}&=&\frac{1}{{e^{\psi}(1+f_{T})}}\bigg[e^{\psi}\big(\frac{f}{2}+\frac{q^{2}}{8\pi
r^{4}}+\frac{q^{2}}{2r^{4}}f_{T}\big)+\big
(-\frac{\lambda''}{2}+\frac{\psi'}{r}+\frac{\lambda'\psi'}{4}-\frac{\lambda'^{2}}
{4}\big)f_{R}
\\\label{21}
&+&\big(\frac{\lambda'}{2}+\frac{2}{r}\big)f'_{R}-\rho
f_{T}e^{\psi}\bigg],
\\\nonumber
P_{t}&=&\frac{1}{{e^{\psi}(1+f_{T})}}\bigg[e^{\psi}\big(\frac{f}{2}-\frac{q^{2}}{8\pi
r^{4}}-\frac{q^{2}}{2r^{4}}f_{T}\big)+\big
(\frac{(\psi'-\lambda')r}{2}+e^{\psi}-1\big)\frac{f_{R}}{r^{2}}
\\\label{22}
&+&\big
(\frac{\lambda'-\psi'}{2}+\frac{1}{r}\big)f'_{R}+f''_{R}-\rho
f_{T}e^{\psi}\bigg].
\end{eqnarray}
The multivariate functions and their derivatives make the field
equations (\ref{20})-(\ref{22}) more complicated.

In order to solve these equations, we assume a specific $f(R,T)$
model as \cite{6}
\begin{equation}\label{23}
f(R,T)=f_{1}(R)+f_{2}(T).
\end{equation}
Different forms of $f_{1}(R)$ with $f_{2}(T)=\gamma T$ \cite{6} are
used to examine the various viable models of curvature matter
coupled gravity. The field equations of model (\ref{23}) become
\begin{eqnarray}\nonumber
\rho&=&\frac{1}{{e^{\psi}}2(2\gamma+1)(\gamma+1)}\bigg[(5\gamma+2)
\bigg\{-e^{\psi}\big(\frac{f}{2}+\frac{q^{2}}{8\pi
r^{4}}+\frac{q^{2}}{2r^{4}}\gamma\big)
\\\nonumber
&+&\big(\frac{\lambda'}{r}-\frac{\lambda'\psi'}{4}
+\frac{\lambda''}{2}
+\frac{\lambda'^{2}}{4}\big)f_{R}+\big(\frac{\psi'}{2}-\frac{2}{r}\big)
f'_{R}-f''_{R}\bigg\}+\gamma\bigg\{e^{\psi}\big(\frac{f}{2}
\\\nonumber
&+&\frac{q^{2}}{8\pi
r^{4}}+\frac{q^{2}}{2r^{4}}\gamma\big)+\big(-\frac
{\lambda''}{2}+\frac{\psi'}{r}+\frac{\lambda'\psi'}{4}-\frac{\lambda'^{2}}{4}\big)f_{R}
+\big(\frac{\lambda'}{2}+\frac{2}{r}\big)f'_{R}\bigg\}
\\\nonumber
&+&2\gamma \bigg\{e^{\psi}\big(\frac{f}{2}-\frac{q^{2}}{8\pi
r^{4}}-\frac{q^{2}}{2r^{4}}\gamma\big)-\big(\frac{(\lambda'-\psi')r}{2}-e^{\psi}+1\big)
\frac{f_{R}}{r^{2}}
\\\label{24}
&+&\big(\frac{\lambda'-\psi'}{2}+\frac{1}{r}\big)
f'_{R}+f''_{R}\bigg\}\bigg],
\\\nonumber
P_{r}&=&\frac{1}{{e^{\psi}}2(2\gamma+1)(\gamma+1)}\bigg[-\gamma\bigg
\{-e^{\psi}\big(\frac{f}{2}+\frac{q^{2}}{8\pi
r^{4}}+\frac{q^{2}}{2r^{4}}\gamma\big)+\big(\frac{\lambda'}{r}
\\\nonumber
&-&\frac{\lambda'\psi'}{4} +\frac{\lambda''}{2}
+\frac{\lambda'^{2}}{4}\big)f_{R}+\big(\frac{\psi'}{2}-\frac{2}{r}\big)
f'_{R}-f''_{R}\bigg\}+(3\gamma+2)\bigg\{e^{\psi}\big(\frac{f}{2}
\\\nonumber
&+&\frac{q^{2}}{8\pi
r^{4}}+\frac{q^{2}}{2r^{4}}\gamma\big)+\big(-\frac
{\lambda''}{2}+\frac{\psi'}{r}+\frac{\lambda'\psi'}{4}-\frac{\lambda'^{2}}{4}\big)f_{R}
+\big(\frac{\lambda'}{2}+\frac{2}{r}\big)f'_{R}\bigg\}
\\\nonumber
&-&2\gamma\bigg\{e^{\psi}\big(\frac{f}{2}-\frac{q^{2}}{8\pi
r^{4}}-\frac{q^{2}}{2r^{4}}\gamma\big)+\big(\frac{(\lambda'-\psi')r}{2}
-e^{\psi}+1\big) \frac{-f_{R}}{r^{2}}
\\\label{25}
&+&\big(\frac{\lambda'-\psi'}{2}+\frac{1}{r}\big)f'_{R}
+f''_{R}\bigg\}\bigg],
\\\nonumber
P_{t}&=&\frac{1}{{e^{\psi}}2(2\gamma+1)(\gamma+1)}\bigg[-\gamma\bigg
\{-e^{\psi}\big(\frac{f}{2}+\frac{q^{2}}{8\pi
r^{4}}+\frac{q^{2}}{2r^{4}}\gamma\big)+\big(\frac{\lambda'}{r}
\\\nonumber
&-&\frac{\lambda'\psi'}{4}+\frac
{\lambda''}{2}+\frac{\lambda'^{2}}{4}\big)f_{R}
+\big(\frac{\psi'}{2}-\frac{2}{r}\big)f'_{R}-f''_{R}\bigg\}+\gamma
\bigg\{e^{\psi}\big(\frac{f}{2}+\frac{q^{2}}{8\pi r^{4}}
\\\nonumber
&+&\frac{q^{2}}{2r^{4}}\gamma\big)+\big(-\frac{\lambda''}{2}+\frac{\psi'}{r}+\frac
{\lambda'\psi'}{4}-\frac{\lambda'^{2}}{4}\big)f_{R}
+\big(\frac{\lambda'}{2}+\frac{2}{r}\big)f'_{R}\bigg\}
\\\nonumber
&-&2(\gamma+1) \bigg\{e^{\psi}\big(\frac{f}{2}-\frac{q^{2}}{8\pi
r^{4}}-\frac{q^{2}}{2r^{4}}\gamma\big)-\big(\frac{(\lambda'-\psi')r}{2}-e^{\psi}+1\big)
\\\label{26}
&\times&\frac{f_{R}}{r^{2}}+\big(\frac{\lambda'-\psi'}{2}+\frac{1}{r}\big)
f'_{R}+f''_{R}\bigg\}\bigg].
\end{eqnarray}
The energy bounds are a set of conditions that are placed on the
behavior of matter and energy in spacetime. In modified theories, if
the null energy condition $((\rho+P_{r}\geq0$),
($\rho+P_{t}+\frac{q^{2}}{4\pi r^{4}}\geq0))$ is violated, then it
shows that the exotic matter is present in the vicinity of the WH
and gives viable traversable WH geometry. This is because the matter
must counteract the gravitational attraction of the WH itself. The
null energy condition has important implications for the viable
traversable WH structure.

In the following, different models of $f(R,T)$ gravity with respect
to $f_{1}(R)$ are analyzed.

\subsection{Model 1}

The exponential model was first proposed by Cognola et al \cite{51}
as
\begin{equation}\label{27}
f(R)=R-LC\left(1-e^{\frac{-R}{C}}\right),
\end{equation}
where $L$ and $C$ are arbitrary constants. This exponential model
provides a powerful framework for understanding both early as well
as late-time cosmic evolution. This model predicts that the
expansion rate of the universe should increase over time, which is
consistent with the observed acceleration. The field equations with
respect to model 1 become
\begin{eqnarray}\nonumber
\rho&=&\frac{1}{{e^{\psi}}2(2\gamma+1)(\gamma+1)}\bigg[(5\gamma+2)\bigg
\{-e^{\psi}\big(\frac{f}{2}+\frac{q^{2}}{8\pi
r^{4}}+\frac{q^{2}}{2r^{4}}\gamma\big)
\\\nonumber
&+&\big(\frac{\lambda'}{r}-\frac{\lambda'\psi'}{4}+\frac
{\lambda''}{2}+\frac{\lambda'^{2}}{4}\big)\big(1-L
e^{\frac{-R}{C}}\big)+\big(\frac{\psi'}{2}-\frac{2}{r}\big)\big(\frac{1}{C}L
e^{\frac{-R}{C}}\big)R'
\\\nonumber
&-&\big\{\big(\frac{1}{C}L
e^{\frac{-R}{C}}\big)R''-\big(\frac{1}{C^{2}}L
e^{\frac{-R}{C}}\big)R'^{2}\big\}\bigg\}+\gamma
\bigg\{e^{\psi}\big(\frac{f}{2}+\frac{q^{2}}{8\pi
r^{4}}+\frac{q^{2}}{2r^{4}}\gamma\big)
\\\nonumber
&+&\big(-\frac{\lambda''}{2}+\frac{\psi'}{r}+\frac
{\lambda'\psi'}{4}-\frac{\lambda'^{2}}{4}\big)(1-Le^{\frac{-R}{C}})+\big
(\frac{\lambda'}{2}+\frac{2}{r}\big)\big(\frac{1}{C}Le^{\frac{-R}{C}}\big)R'\bigg\}
\\\nonumber
&+&2\gamma \bigg\{e^{\psi}\big(\frac{f}{2}-\frac{q^{2}}{8\pi
r^{4}}-\frac{q^{2}}{2r^{4}}\gamma\big)+\big(\frac{(\lambda'-\psi')r}{2}-e^{\psi}+1\big)
\frac{(-1+L e^{\frac{-R}{C}})}{r^{2}}
\\\nonumber
&+&\big(\frac{\lambda'-\psi'}{2}+\frac{1}{r}\big)\big(\frac{1}{C}L
e^{\frac{-R}{C}}\big)R'+\big\{\big(\frac{1}{C}L
e^{\frac{-R}{C}}\big)R''
\\\label{28}
&+&\big(\frac{-1}{C^{2}}L
e^{\frac{-R}{C}}\big)R'^{2}\big\}\bigg\}\bigg],
\\\nonumber
P_{r}&=&\frac{1}{{e^{\psi}}2(2\gamma+1)(\gamma+1)}\bigg[-\gamma\bigg
\{-e^{\psi}\big(\frac{f}{2}+\frac{q^{2}}{8\pi
r^{4}}+\frac{q^{2}}{2r^{4}}\gamma\big)+\big(\frac{\lambda'}{r}
\\\nonumber
&-&\frac{\lambda'\psi'}{4}+\frac
{\lambda''}{2}+\frac{\lambda'^{2}}{4}\big)\big(1-L
e^{\frac{-R}{C}}\big)+\big(\frac{\psi'}{2}-\frac{2}{r}\big)\big(\frac{1}{C}L
e^{\frac{-R}{C}}\big)R'
\\\nonumber
&-&\big\{\big(\frac{1}{C}L
e^{\frac{-R}{C}}\big)R''+\big(\frac{-1}{C^{2}}L
e^{\frac{-R}{C}}\big)R'^{2}\big\}\bigg\}+(3\gamma+2)
\bigg\{e^{\psi}\big(\frac{f}{2}+\frac{q^{2}}{8\pi r^{4}}
\\\nonumber
&+&\frac{q^{2}}{2r^{4}}\gamma\big)+\big(-\frac{\lambda''}{2}+\frac{\psi'}{r}+\frac
{\lambda'\psi'}{4}-\frac{\lambda'^{2}}{4}\big)\big(1-L
e^{\frac{-R}{C}}\big)+\big(\frac{\lambda'}{2}
\\\nonumber
&+&\frac{2}{r}\big)\big(\frac{1}{C}L
e^{\frac{-R}{C}}\big)R'\bigg\}-2\gamma\bigg
\{e^{\psi}\big(\frac{f}{2}-\frac{q^{2}}{8\pi
r^{4}}-\frac{q^{2}}{2r^{4}}\gamma\big)-\big(\frac{(\lambda'-\psi')r}{2}
\\\nonumber
&-&e^{\psi}+1\big) \frac{\big(1-L
e^{\frac{-R}{C}}\big)}{r^{2}}+\big(\frac{\lambda'-\psi'}{2}+\frac{1}{r}\big)\big(\frac{1}{C}L
e^{\frac{-R}{C}}\big)R'+\big\{\big(\frac{1}{C}L
e^{\frac{-R}{C}}\big)
\\\label{29}
&\times&R''+\big(\frac{-1}{C^{2}}L
e^{\frac{-R}{C}}\big)R'^{2}\big\}\bigg\}\bigg],
\\\nonumber
P_{t}&=&\frac{1}{{e^{\psi}}2(2\gamma+1)(\gamma+1)}\bigg[-\gamma\bigg
\{-e^{\psi}\big(\frac{f}{2}+\frac{q^{2}}{8\pi
r^{4}}+\frac{q^{2}}{2r^{4}}\gamma\big)+\big(\frac{\lambda'}{r}
\\\nonumber
&-&\frac{\lambda'\psi'}
{4}+\frac{\lambda''}{2}+\frac{\lambda'^{2}}{4}\big)\big(1-L
e^{\frac{-R}{C}}\big)+\big(\frac{\psi'}{2}-\frac{2}{r}\big)\big(\frac{1}{C}L
e^{\frac{-R}{C}}\big)R'
\\\nonumber
&-&\big\{\big(\frac{1}{C}L
e^{\frac{-R}{C}}\big)R''+\big(\frac{-1}{C^{2}}L
e^{\frac{-R}{C}}\big)R'^{2}\big\}\bigg\}+\gamma
\bigg\{e^{\psi}\big(\frac{f}{2}+\frac{q^{2}}{8\pi
r^{4}}+\frac{q^{2}}{2r^{4}}\gamma\big)
\\\nonumber
&+&\big(-\frac{\lambda''}{2}+\frac{\psi'}{r}+\frac
{\lambda'\psi'}{4}-\frac{\lambda'^{2}}{4}\big)\big(1-L
e^{\frac{-R}{C}}\big)
+\big(\frac{\lambda'}{2}+\frac{2}{r}\big)\big(\frac{1}{C}L
e^{\frac{-R}{C}}\big)R'\bigg\}
\\\nonumber
&-&2(\gamma+1)\bigg\{e^{\psi}\big(\frac{f}{2}-\frac{q^{2}}{8\pi
r^{4}}-\frac{q^{2}}{2r^{4}}\gamma\big)-\big(\frac{(\lambda'-\psi')r}{2}-e^{\psi}+1\big)
\\\nonumber
&\times&\frac{\big(1-L e^{\frac{-R}{C}}\big)}{r^{2}}+\big
(\frac{\lambda'-\psi'}{2}+\frac{1}{r}\big)\big(\frac{1}{C}L
e^{\frac{-R}{C}}\big)R'+\big\{\big(\frac{1}{C}L
e^{\frac{-R}{C}}\big)R''
\\\label{30}
&+&\big(\frac{-1}{C^{2}}L
e^{\frac{-R}{C}}\big)R'^{2}\big\}\bigg\}\bigg].
\end{eqnarray}
We consider radial dependent form of the charge as \cite{52,53}
\begin{eqnarray}{\nonumber}
q(r)=\beta r^{3},
\end{eqnarray}
where $\beta$ is an arbitrary constant. We choose $\beta=0.01$ for
our convenience in all the graphs. The graphical behavior of null
energy condition in the presence of charge is shown in Figure
\textbf{2}. From Figure \textbf{2}, we can see that for $L=0.5$,
$C=-0.5$, (black color), $L=0.7$, $C=-1$ (green color), $L=0.8$,
$C=-2$ (red color), the null energy condition is not satisfied.
Hence the WH vicinity is filled with exotic matter and consequently,
we obtain viable traversable WH structure.
\begin{figure}
\epsfig{file=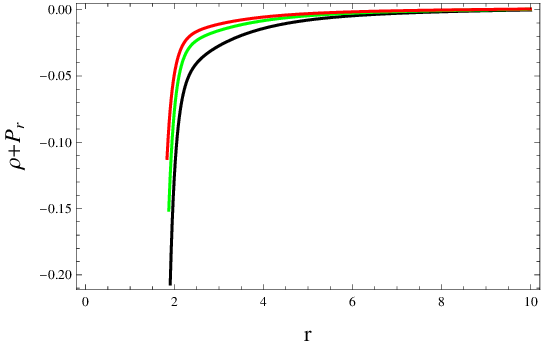,width=.5\linewidth}
\epsfig{file=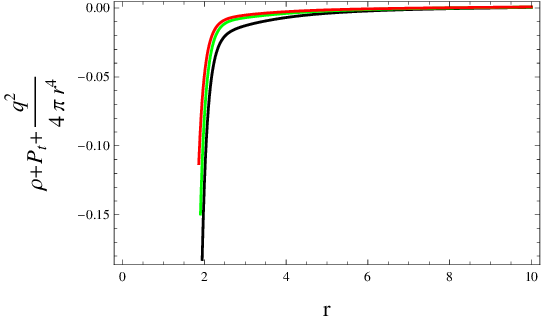,width=.5\linewidth}\caption{Plots of null energy
condition versus $r$ with respect to model 1.}
\end{figure}

\subsection{Model 2}

The Starobinsky model is another well-known gravity model consistent
with cosmic observations and fulfills the solar system tests. The
Starobinsky model is defined as \cite{54}
\begin{equation}\label{31}
f(R)=R-MC\bigg[1-\bigg(1+\frac{R^{2}}{C^{2}}\bigg)^{-p}\bigg].
\end{equation}
This model predicts a particular pattern of gravitational waves and
can also reproduce the early cosmic inflation. During this phase,
the universe expanded exponentially, which can explain several
observed features of the cosmic microwave background radiation.
Here, $M$ and $p$ are arbitrary constants. The field equations with
respect to the Starobinsky model are
\begin{eqnarray}\nonumber
\rho&=&\frac{1}{{e^{\psi}}2(2\gamma+1)(\gamma+1)}\bigg[(5\gamma+2)
\frac{1}{e^{\psi}}\bigg\{-e^{\psi}\big(\frac{f}{2}+\frac{q^{2}}{8\pi
r^{4}}+\frac{q^{2}}{2r^{4}}\gamma\big)
\\\nonumber
&+&\big(\frac{\lambda'}{r}-
\frac{\lambda'\psi'}{4}+\frac{\lambda''}{2}+\frac{\lambda'^{2}}{4}\big)\big(1-\frac{2pMR
(1+\frac{R^{2}}{C^{2}})^{-1-p}}{C}\big)
\\\nonumber
&+&\big(\frac{\psi'}{2}-\frac{2}{r}\big)\bigg\{\frac{2Mp}{C(1+\frac
{R^{2}}{C^{2}})^{1+p}}\bigg(\frac{-2(1+p)R^{2}}{C^{2}(1+\frac{R^{2}}{C^{2}})}-1\bigg)
R'\bigg\}
\\\nonumber
&-&\bigg\{\frac{2Mp}{C(1+\frac{R^{2}}{C^{2}})^{1+p}}\bigg
(\frac{-2(1+p)R^{2}}{C^{2}(1+\frac{R^{2}}{C^{2}})}-1\bigg)R''+\frac{4pMR}{C^{3}
(1+\frac{R^{2}}{C^{2}})^{p+2}}
\\\nonumber
&\times&\bigg((3+
3p)-\frac{2(p^{2}+3p+4)R^{2}}{C^{2}(1+\frac{R^{2}}{C^{2}})}\bigg)R'^{2}
\bigg\}\bigg\}+\gamma
\bigg\{e^{\psi}\big(\frac{f}{2}+\frac{q^{2}}{8\pi r^{4}}
\\\nonumber
&+&\frac{q^{2}}{2r^{4}}\gamma\big)+\big(-\frac{\lambda''}{2}+\frac{\psi'}{r}+\frac
{\lambda'\psi'}{4}-\frac{\lambda'^{2}}{4}\big)\big(1-\frac{2pMR(1+\frac
{R^{2}}{C^{2}})^{-1-p}}{C}\big)
\\\nonumber
&+&\big(\frac{\lambda'}{2}+\frac{2}{r}\big)\bigg\{\frac{2Mp}{C(1
+\frac{R^{2}}{C^{2}})^{1+p}}\bigg(\frac{-2(1+p)R^{2}}{C^{2}(1+\frac
{R^{2}}{C^{2}})}-1\bigg)R'\bigg\}\bigg\}
\\\nonumber
&+&2\gamma\bigg\{e^{\psi}\big(\frac{f}{2}-\frac{q^{2}}{8\pi
r^{4}}-\frac{q^{2}}{2r^{4}}\gamma\big)+\big(\frac{(\lambda'-\psi')r}{2}-e^{\psi}+1\big)
\\\nonumber
&\times&\frac{\big(-1+\frac
{2pMR(1+\frac{R^{2}}{C^{2}})^{-1-p}}{C}\big)}{r^{2}}+\bigg(\frac
{\lambda'-\psi'}{2}+\frac{1}{r}\big)\bigg\{\frac{2Mp}{C(1+\frac{R^{2}}{C^{2}})^{1+p}}
\\\nonumber
&\times&\bigg
(\frac{-2(1+p)R^{2}}{C^{2}(1+\frac{R^{2}}{C^{2}})}-1\bigg)R'\bigg\}
+\bigg\{\frac{2Mp}{C(1+\frac{R^{2}}{C^{2}})^{1+p}}
\\\nonumber
&\times&\bigg(\frac{-2(1+p)R^{2}}{C^{2}(1+\frac{R^{2}}{C^{2}})}-1\bigg)
R''+\frac{4pMR}{C^{3}(1+\frac{R^{2}}{C^{2}})^{p+2}}\bigg(3(1+p)
\\\label{32}
&-&\frac{2(p^{2}+3p+4)R^{2}}{C^{2}(1+\frac{R^{2}}{C^{2}})}
\bigg)R'^{2}\bigg\}\bigg\}\bigg],
\\\nonumber
P_{r}&=&\frac{1}{{e^{\psi}}2(2\gamma+1)(\gamma+1)}\bigg[-\gamma\bigg
\{-e^{\psi}\big(\frac{f}{2}+\frac{q^{2}}{8\pi
r^{4}}+\frac{q^{2}}{2r^{4}}\gamma\big)
\\\nonumber
&+&\big(\frac{\lambda'}{r}-\frac{\lambda'\psi'}{4}+\frac
{\lambda''}{2}+\frac{\lambda'^{2}}{4}\big)\big(1-\frac{2pMR
(1+\frac{R^{2}}{C^{2}})^{-1-p}}{C}\big)+\big(\frac{\psi'}{2}
\\\nonumber
&-&\frac{2}{r}
\big)\bigg\{\frac{2Mp}{C(1+\frac{R^{2}}{C^{2}})^{1+p}}\bigg(\frac{
-2(1+p)R^{2}}{C^{2}(1+\frac{R^{2}}{C^{2}})}-1\bigg)R'\bigg\}-\bigg\{\frac{2Mp}{C(1+\frac{R^{2}}{C^{2}})^{1+p}}
\\\nonumber
&\times&\bigg(\frac{-2(1+p)R^{2}}{C^{2}(1+\frac{R^{2}}{C^{2}})}-1\bigg)R''+\frac
{4nMR}{C^{3}(1+\frac{R^{2}}{C^{2}})^{p+2}}\bigg(3(1+p)
\\\nonumber
&-&\frac{2(p^{2}+3p+4)R^{2}}{C^{2}(1+\frac{R^{2}}{C^{2}
})}\bigg)R'^{2}\bigg\}\bigg\}+(3\gamma+2)\bigg\{e^{\psi}\big(\frac{f}{2}+\frac{q^{2}}{8\pi
r^{4}}
\\\nonumber
&+&\frac{q^{2}}{2r^{4}}\gamma\big)+\big
(-\frac{\lambda''}{2}+\frac{\psi'}{r}+\frac{\lambda'\psi'}
{4}-\frac{\lambda'^{2}}{4}\big)\big(1-\frac{2pMR(1+\frac{R^{2}}{C^{2}})
^{-1-p}}{C}\big)
\\\nonumber
&+&\big(\frac{\lambda'}{2}+\frac{2}{r}\big)\bigg\{\frac{2Mp}
{C(1+\frac{R^{2}}{C^{2}})^{1+p}}\bigg(\frac{-2(1+p)R^{2}}{C^{2}(1+\frac{R^{2}}{C^{2}})}-1\bigg)
R'\bigg\}\bigg\}
\\\nonumber
&-&2\gamma\bigg\{e^{\psi}\big(\frac{f}{2}-\frac{q^{2}}{8\pi
r^{4}}-\frac{q^{2}}{2r^{4}}\gamma\big)+\big(\frac{(\lambda'-\psi')
r}{2}-e^{\psi}+1\big)
\\\nonumber
&\times&\frac{\big(-1+\frac{2pMR(1+\frac{R^{2}}{C^{2}})^{-1-p}}{C}
\big)}{r^{2}}+\big(\frac{\lambda'-\psi'}{2}+\frac{1}{r}\big)\bigg\{\frac
{2Mp}{C(1+\frac{R^{2}}{C^{2}})^{1+p}}
\\\nonumber
&\times&\bigg(\frac{-2(1+p)R^{2}}{C^{2}(1+\frac{R^{2}}{C^{2}})}-1\bigg)R'
\bigg\}+\bigg\{\frac{2Mp}{C(1+\frac{R^{2}}{C^{2}})^{1+p}}
\\\nonumber
&\times&\bigg(\frac{
-2(1+p)R^{2}}{C^{2}(1+\frac{R^{2}}{C^{2}})}-1\bigg)R''
+\frac{4pMR}{C^{3}(1+\frac{R^{2}}{C^{2}})^{p+2}}\bigg(3(1+p)
\\\label{33}
&-&\frac{2(p^{2}+3p+4)R^{2}}{C^{2}(1+\frac{R^{2}}{C^{2}})}\bigg)R'^{2}
\bigg\}\bigg\}\bigg],
\\\nonumber
P_{t}&=&\frac{1}{{e^{\psi}}2(2\gamma+1)(\gamma+1)}\bigg[-\gamma
\bigg\{-e^{\psi}\big(\frac{f}{2}+\frac{q^{2}}{8\pi
r^{4}}+\frac{q^{2}}{2r^{4}}\gamma\big)
\\\nonumber
&+&\big(\frac{\lambda'}{r}-\frac{\lambda'\psi'}
{4}+\frac{\lambda''}{2}+\frac{\lambda'^{2}}{4}\big)\big(1-\frac{2pMR
(1+\frac{R^{2}}{C^{2}})^{-1-p}}{C}\big)+\big(\frac{\psi'}{2}-\frac{2}
{r}\big)
\\\nonumber
&\times&\bigg\{\frac{2Mp}{C(1+\frac{R^{2}}{C^{2}})^{1+p}}\bigg
(\frac{-2(1+p)R^{2}}{C^{2}(1+\frac{R^{2}}{C^{2}})}-1\bigg)R'\bigg\}
-\bigg\{\frac{2Mp}{C(1+\frac{R^{2}}{C^{2}})^{1+p}}
\\\nonumber
&\times&\bigg(\frac{-2(1+p)R^{2}}{C^{2}(1+\frac{R^{2}}{C^{2}})}-1
\bigg)R''+\frac{4pMR}{C^{3}(1+\frac{R^{2}}{C^{2}})^{p+2}}
\bigg(3(1+p)
\\\nonumber
&-&\frac{2(p^{2}+3p+4)R^{2}}{C^{2}(1+\frac{R^{2}}
{C^{2}})}\bigg)R'^{2}\bigg\}\bigg\}+\gamma\bigg\{e^{\psi}\big(\frac{f}{2}+\frac{q^{2}}{8\pi
r^{4}}+\frac{q^{2}}{2r^{4}}\gamma\big)
\\\nonumber
&+&\big(-\frac{\lambda''}{2}+\frac{\psi'}{r}+\frac{\lambda'\psi'}{4}-\frac{\lambda'^{2}}{4}\big)
\big(1-\frac{2pRM(1+\frac{R^{2}}{C^{2}})^{-1-p}}{C}\big)
\\\nonumber
&+&\big(\frac{\lambda'}{2}
+\frac{2}{r}\big)\bigg\{\frac{2Mp}{C(1+\frac{R^{2}}{C^{2}})^{1+p}}
\bigg(\frac{-2(1+p)R^{2}}{C^{2}(1+\frac{R^{2}}{C^{2}})}-1\bigg)
R'\bigg\}\bigg\}
\\\nonumber
&-&2(\gamma+1)\bigg\{e^{\psi}\big(\frac{f}{2}-\frac{q^{2}}{8\pi
r^{4}}-\frac{q^{2}}{2r^{4}}\gamma\big)+\big(\frac{
(\lambda'-\psi')r}{2}-e^{\psi}+1\big)
\\\nonumber
&\times&\frac{\big(-1+\frac{2pMR(1+\frac{R^{2}}{C^{2}})^{-1-p}}
{C}\big)}{r^{2}}+\big(\frac{\lambda'-\psi'}{2}+\frac{1}{r}\big)\bigg
\{\frac{2Mp}{C(1+\frac{R^{2}}{C^{2}})^{1+p}}
\\\nonumber
&\times&\bigg(\frac{-2(1+p)R^{2}}{C^{2}(1+\frac{R^{2}}{C^{2}})}
-1\bigg)R'\bigg\}+\bigg\{\frac{2Mp}{C(1+\frac{R^{2}}{C^{2}})^{1+p}}
\\\nonumber
&\times&\bigg(\frac{-2(1+p)R^{2}}{C^{2}(1+\frac{R^{2}}{C^{2}})}-1\bigg)R''
+\frac{4pMR}{C^{3}(1+\frac{R^{2}}{C^{2}})^{p+2}}\bigg(3(1+p)
\\\label{34}
&-&\frac{2(p^{2}+3p+4)R^{2}}{C^{2}(1+\frac{R^{2}}{C^{2}})}
\bigg)R'^{2}\bigg\}\bigg\}\bigg].
\end{eqnarray}
Figure \textbf{3} shows that the null energy condition is violated
for $p=-7,~M=1,~C=-23$, (black color), $p=-7.5,~M=-1,~C=-22$ (green
color), $p=-8,~M=1.8,~C=-21$ (red color). Hence we obtain
traversable spherically symmetric WH geometry.
\begin{figure}
\epsfig{file=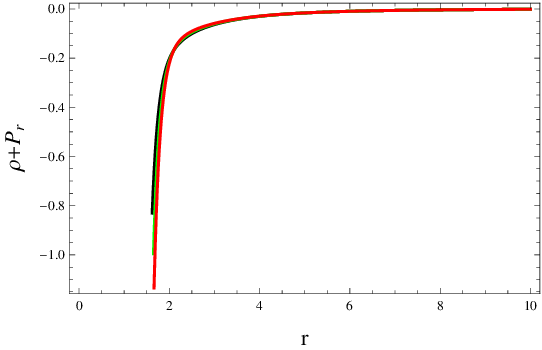,width=.5\linewidth}
\epsfig{file=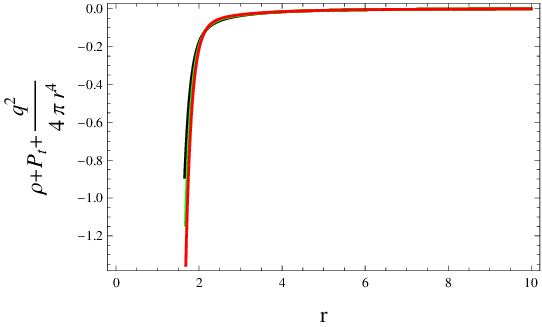,width=.5\linewidth}\caption{Plots of null energy
condition versus $r$ with respect to model 2.}
\end{figure}

\section{Stability Analysis}

Stability is a crucial factor while analyzing cosmic structures such
as star clusters, and even the universe as a whole. The stability of
cosmic structures provides insights into the underlying physics and
processes that govern their behavior. Stable structures are those
that maintain their shape and properties over time, despite any
external influences or disturbances. Studying these objects can help
scientists to better understand the distribution of matter in the
universe, the formation and evolution of galaxies clusters and the
nature of dark components. Here, we use sound speed to check the
stability of our obtained WH geometry.

We use the causality and Herrera cracking techniques to examine the
stability of WH under the influence of charge. The causality
constraint states that no information or signal can propagate faster
than the speed of light. This condition implies that the speed of
sound components $(v^{2}_{r}=\frac{dP_{r}}{d\rho},~v_{t}^{2}
=\frac{dP_{t}}{d\rho})$ must lie between $[0,1]$ for stable cosmic
structure. From Figure \textbf{4}, we can see that the causality
condition is satisfied for our obtained WH solutions under the
influence of charge and dark source terms. Herrera cracking is
another method that considers the WH as a thin-shell and analyzes
the behavior of stress under various perturbations. Herrera cracking
method states that the difference between the sound speed components
$v_{t}^{2},~v^{2}_{r}$ must lie between $0$ to $1$. The graphical
behavior is presented in Figure \textbf{5}, which determines that
our obtained WH solutions satisfies all the necessary criteria of
stability. Hence, physically viable and stable WH solutions exist in
$f(R,T)$ theory.
\begin{figure}
\epsfig{file=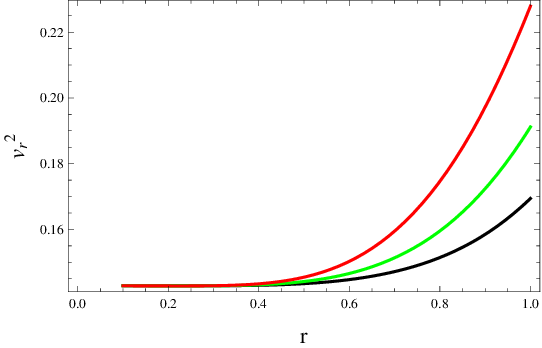,width=.5\linewidth}
\epsfig{file=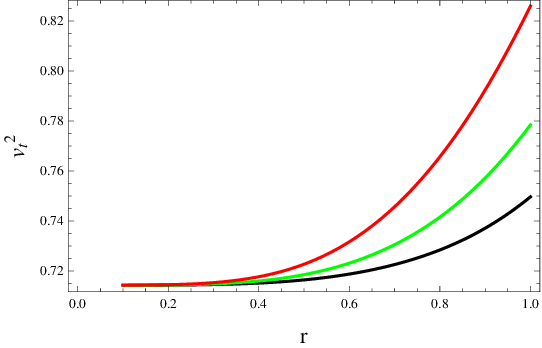,width=.5\linewidth}
\epsfig{file=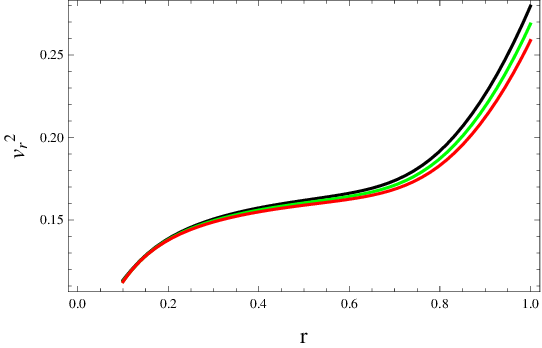,width=.5\linewidth}
\epsfig{file=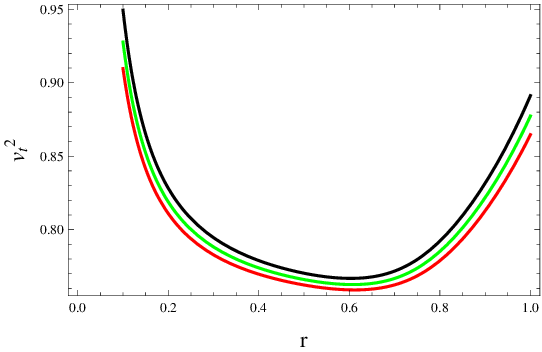,width=.5\linewidth}\caption{Plots of causality
condition versus radial coordinate with respect to models 1 and 2.}
\end{figure}
\begin{figure}
\epsfig{file=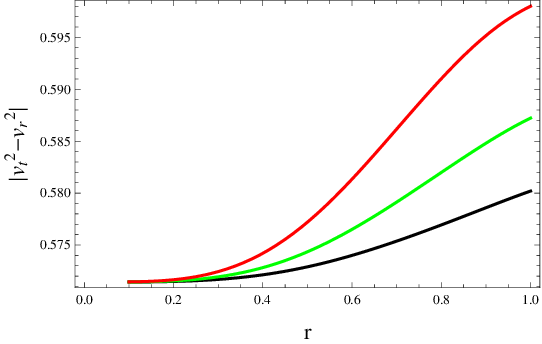,width=.5\linewidth}
\epsfig{file=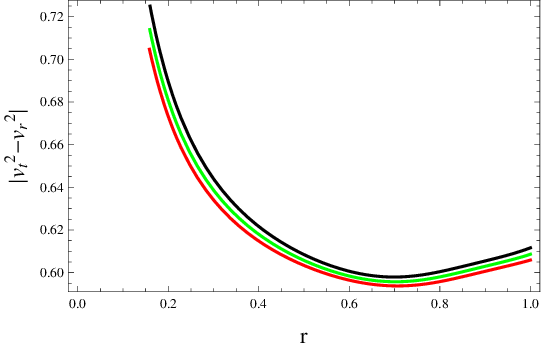,width=.5\linewidth}\caption{Plots of Herrera
cracking versus radial coordinate with respect to models 1 and 2.}
\end{figure}

\section{Final Remarks}

In the literature, there are several approaches that have been used
to find viable WH solutions. One common approach for evaluating the
shape function is to make certain assumptions about the matter
ingredients. These assumptions can be used to explore the Einstein
field equations and derive the shape function for the WH structure.
Another approach is to consider a shape function to examine the
behavior of energy bounds. The energy conditions can be used to
derive important physical properties of the object. In this paper,
we have studied the effects of charge on viable WH geometry in
$f(R,T)$ gravity. For this, we have built a shape function by
employing the Karmarkar condition. We have assumed exponential
gravity and Starobinsky gravity models of this gravity to analyze
the viability and stability of WH solutions. The viability of
obtained WH solutions is analyzed by null energy condition, and
their stable state is checked through the speed of sound. The
obtained results are summarized as follows
\begin{itemize}
\item
We have shown that our developed shape function is physically viable
as it satisfies all the necessary conditions (Figure \textbf{1}).
\item
For the exponential gravity model, the null energy condition is
violated, and exotic matter exists in the WH vicinity for $L>0$ and
$C<0$. Thus we have obtained the viable traversable WH geometry for
specific values of $L$ and $C$.
\item
The null energy condition is violated for $C<0$, $M>0$ and $p<0$
corresponding to Starobinsky gravity model (Figure \textbf{3}). This
manifests that the interior region of WH throat is filled with
exotic matter, which leads to the viable traversable WH geometry.
\item
We have found that all the required constraints of stability are
fulfilled under the influence of charge in $f(R,T)$ theory (Figures
\textbf{4} and \textbf{5}).
\end{itemize}
It is important to note here that our obtained solutions are more
stable in the presence of charge as compared to that found in the
absence of electromagnetic field \cite{46,47}.

\end{document}